\documentclass[epj]{svjour}
\usepackage{graphics,epsfig}
\begin{document}
\title{Immunization of Real Complex Communication Networks}

\author{Jes\'us G\'omez-Garde\~nes\inst{1,2} \and Pablo Echenique\inst{1,3}
\and Yamir Moreno\inst{1,3}
}                     
\offprints{Y.Moreno (yamir@unizar.es)}          
\institute{Institute for Biocomputation and Physics of Complex
Systems (BIFI), University of Zaragoza, Zaragoza 50009, Spain \and Departamento de F\'{\i}sica de la Materia Condensada,
Universidad de Zaragoza, Zaragoza 50009, Spain \and Departamento de F\'{\i}sica Te\'orica, Universidad de
Zaragoza, Zaragoza 50009, Spain}
\date{Received: date / Revised version: date}
%
\abstract{Most communication networks are complex. In this paper, we address
  one of the fundamental problems we are facing nowadays, namely, how
  we can efficiently protect these networks. To this end, we study an
  immunization strategy and found that it works almost as good as targeted
  immunization, but using only local information about the network
  topology. Our findings are supported with numerical simulations of
  the Susceptible-Infected-Removed (SIR) model on top of real
  communication networks, where immune nodes are previously identified
  by a covering algorithm. The results provide useful hints in the way
  to design and deploying a digital immune system.
} 
\maketitle

\section{Introduction}

Communications networks have been intensively studied during the last several
years as it turned out that their topology is far from being random
\cite{fff99,pvv01,vazquez-2002,pv04}. In particular, it has been found that
physical networks $-$the Internet$-$ as well as logical$-$ the World Wide
Web$-$ and peer-to-peer networks $-$Gnutella$-$ are characterized by a power
law degree distribution \cite{pv04} (thus, they are referred to as scale-free
networks \cite{ba99,barabasi-1999}), $P(k)\sim k^{-\gamma}$, where the degree or
connectivity $k$ of a node is the number of nodes it is attached to. These
findings, together with similar network structures found in fields as diverse
as biological, social and natural systems
\cite{dorogovtsev-2003-book,newman03a,book2}, have led to a burst of activity
aimed at characterizing the structure and dynamics of complex networks.

The spreading of an epidemic disease in complex networks was among the
relevant problems that were first addressed in the literature
\cite{ceah00,pv00,pastor-satorras-2001}. Surprisingly, it was found that for
infinite scale-free networks with $2<\gamma<3$, the epidemic always
pervades the system no matter what the spreading rate is
\cite{pv00,pastor-satorras-2001,mpv02,lm01a}, even when correlations
are taken into account \cite{boguna-2003a,moreno-2003,vm02}. In other
words, the usual threshold picture does not apply anymore. This fact
would be a mere anecdote if not because most vaccination and public
health campaigns are based on the existence of such a threshold
\cite{am92}. In practice, it would be desirable to have a threshold as
large as possible for a given epidemic disease.

Soon after the first studies on epidemic spreading, it was realized
that traditional vaccination strategies based on random immunization,
while worth taking for random network topologies, were useless in
scale-free networks\cite{pv01c}. Specifically, it was shown that a
minimum fraction as large as $80\%$ of the nodes has to be immunized
in order to recover the epidemic threshold. New vaccination strategies
are thus needed in order to efficiently deal with the actual topology
of real-world networks. A very efficient approach consists of
vaccinating the highly connected nodes in order to cut the path
through which most of the susceptible nodes catch the epidemics
\cite{pv01c,db01a}. However, in order to do that, one has to identify
the core groups or hubs of the system. In general, this is extremely
unrealistic, particularly for large networks and systems lacking
central organizational rules such as social networks.

In this paper, we consider the immunization problem from a different
perspective. We show that it can be treated as a covering problem, in
which a set of immune agents has to be placed somewhere in the
network.  The main advantage of this approach is that only local
topological knowledge is needed up to a given distance $d$, so that it
can be straightforwardly applied to a real situation. To verify the
results of the immunization strategy, we implement the
Susceptible-Infected-Removed epidemiological model \cite{mpv02,lm01a}
on top of the Internet maps at the Autonomous Systems (AS) and router
levels \cite{pvv01,vazquez-2002,pv04} and compare with the results
obtained by using targeted and random immunization as well as a local
immunization strategy. Our results indicate that the algorithm
performs quite well and is near the optimal one. On the other hand, we
show that the efficiency of the vaccination strongly depends on the
degree-degree correlations as the covering outcome is directly related
to the structure of the underlying network.

\section{Susceptible-Infected-Removed model on Real Nets}

In order to be able to compare the efficiency of the different
immunization strategies, we first perform extensive numerical
simulations of an epidemic spreading process on top of real
architectures (here, epidemics refers to any undesired spreading
process, i.e, virus, spam, etc). We consider the SIR model as a
plausible model for epidemic spreading \cite{am92,mpv02}. In this
model, nodes can be in three different states.  Susceptible nodes,
$S$, have not been infected and are healthy. They catch the disease
via direct contact with infected nodes, $I$, at a rate $\lambda$.
Finally, recovered nodes, $R$, are those nodes that have caught the
disease and have stopped spreading it with probability $\beta$
(without loss of generality, $\beta$ has been set to 1
henceforth). The relevant order parameter of the disease dynamics is
the total number of nodes (or the fraction of them, $R$) that got
infected once the epidemic process dies out, i.e., when no infected
nodes are left in the system.

\begin{figure}
\begin{center}
\epsfig{file=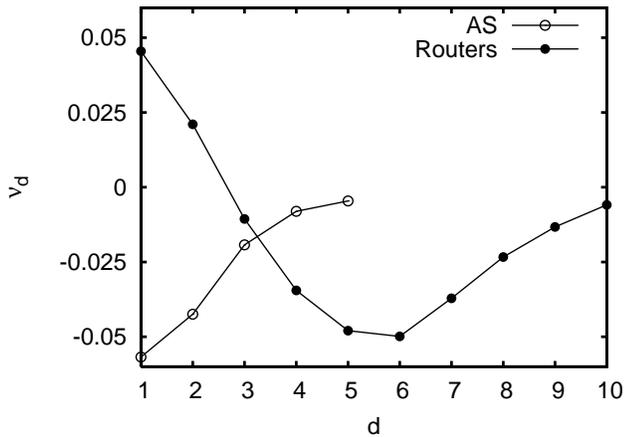,width=2.4in,angle=-90,clip=1}
\end{center} 
\caption{Correlations as a function of $d$ for the AS and router
  graph representations of the Internet. $\nu_d$ is the slope of the
  curve $\langle K^{(d)} \rangle _k$, which measures the average
  degree of neighbors at a distance $d$. See \cite{pablo1} for details
  of this quantity.}
\label{fignu} 
\end{figure}

\begin{figure}
\begin{center}
\epsfig{file=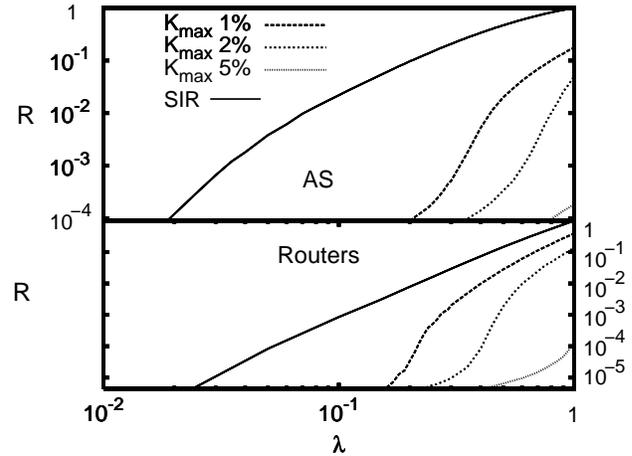,width=3.3in,angle=0,clip=1}
\end{center} 
\caption{Final fraction of infected nodes for the SIR model and targeted
  immunization with different number of immunized nodes for the AS (a) and
  router (b) map representations of the Internet. Simulations were carried out
  starting from a single infected node at $t=0$ in all cases. The plots are in
  a log-log scale for a better visualization.}
\label{fig1} 
\end{figure}

On the other hand, the simulations performed throughout this work have
been carried out on real communication networks. The fact that any
study thought to have practical applications should be tested in real
systems led us to such an election. These networks have unique
topological properties difficult to gather with existing generic
network models $-$ namely, degree-degree correlations and clustering
properties. The networks on top of which numerical simulations of the
immunization strategies and the SIR dynamics have been performed are
the following. AS: Autonomous system level graph representation of the
Internet as of April 16th, 2001 \cite{asref}. Gnutella: Snapshot of
the Gnutella peer to peer network, provided by Clip2 Distributed
Search Solutions. Router: Router level graph representation of the
Internet \cite{routerref}.  The three networks are sparse and show an
average degree around 3.  Additionally, they are small-worlds
\cite{ws98} with an average distance between vertices less than 10,
and they are characterized by a power law degree distribution
$P(k)\sim k^{-\gamma}$, with $\gamma\approx2.2$. A detailed
characterization of these graphs is presented in Refs.
\cite{vazquez03} (Gnutella) and \cite{pvv01,pv04,vpv02b} (AS and
Router graphs).

These networks share a number of topological features but are
radically different in their degree-degree correlations. Correlations
are usually defined taking into account the degrees of
nearest-neighbors. We have recently shown \cite{pablo1}, however, that
whether a network can be regarded as assortative (when correlations
are positive, i.e., there is a tendency to establish connections
between vertices with similar degrees) or disassortative (negative
correlations, the tendency is the opposite) depends on the distance
used to average the degrees of the neighboring vertices. The AS and
the Gnutella graphs show dissortative correlations for any value of
$d$, though the correlations are smoothed as $d$ grows. On the other
hand, in the Router network, the degree correlations are assortative
up to $d=2$. However, for $d>2$ the correlations become dissortative
and beyond $d>6$ start to approach the uncorrelated limit as shown in
Fig.\ \ref{fignu} \cite{pablo1}. These peculiar properties directly
affect the outcome of algorithms run on top of these networks.

In the following, we focus on the results obtained for the AS and
router maps of the Internet. The behavior of both the epidemic
spreading process and the immunization strategies for the Gnutella
graph are qualitatively the same as for the AS map, with the only
difference of more pronounced finite-size effects.

We have performed Monte Carlo simulations of the SIR model on top of the
Internet maps. Starting from an initial state in which a randomly chosen node
is infected, susceptible nodes catch the disease (or virus) with probability
$\lambda$ if they contact a spreader. In its turn, infected vertices become
removed and do not take part anymore in the spreading process at a rate
$\beta=1$. The fraction of removed nodes, $R$, when no spreaders are left in
the system gives the epidemic incidence. All results have been averaged over at
least 1000 realizations corresponding to different initially infected nodes.
Figure\ \ref{fig1} shows the epidemic incidence in the AS and router maps of
the Internet as a function of the spreading rate $\lambda$.

As can be seen from the figure, the epidemic threshold is slightly
larger in the router graph than in the network made up of AS's. This
difference in the behavior of the SIR model on different
representations of the Internet may be understood from the distinct
degree-degree correlations shown by both graphs. Though we think that
the main differences in the algorithm's performance are due to
correlations, it should be noticed that a number of other topological
features such as clustering and hierarchy properties may also be at
the root of the different behaviors. Our guess is mainly based on the
performance of local algorithms such as the covering recipe that we
will use in the next section. As for correlations, in the AS map
representation, highly connected nodes are likely connected to nodes
with smaller degrees. Therefore, the spreading process generally
passes alternatively from highly to poorly connected nodes. In this
way, the epidemics has more chances to reach a number of nodes other
than the hubs. This is not the case of the Router map, where it is
more likely that hubs are grouped together and that once one of them
get infected, its neighbors (also highly connected nodes) do
so. However, when the epidemics leaves the hubs, the remaining
(uninfected) nodes are, likely, poorly connected and with high
probability the process will die out, specially for small values of
$\lambda\sim\lambda_c$. That is, in the router map, the epidemic
reaches the hubs, but then goes down to nodes of decreasing degree and
stops soon afterwards, resulting in a smaller fraction of infected
nodes (the hubs and a few more, i.e, a tiny fraction of the network)
and thus to an effective threshold that is larger than that for the
AS.

\begin{figure} 
\begin{center}
\epsfig{file=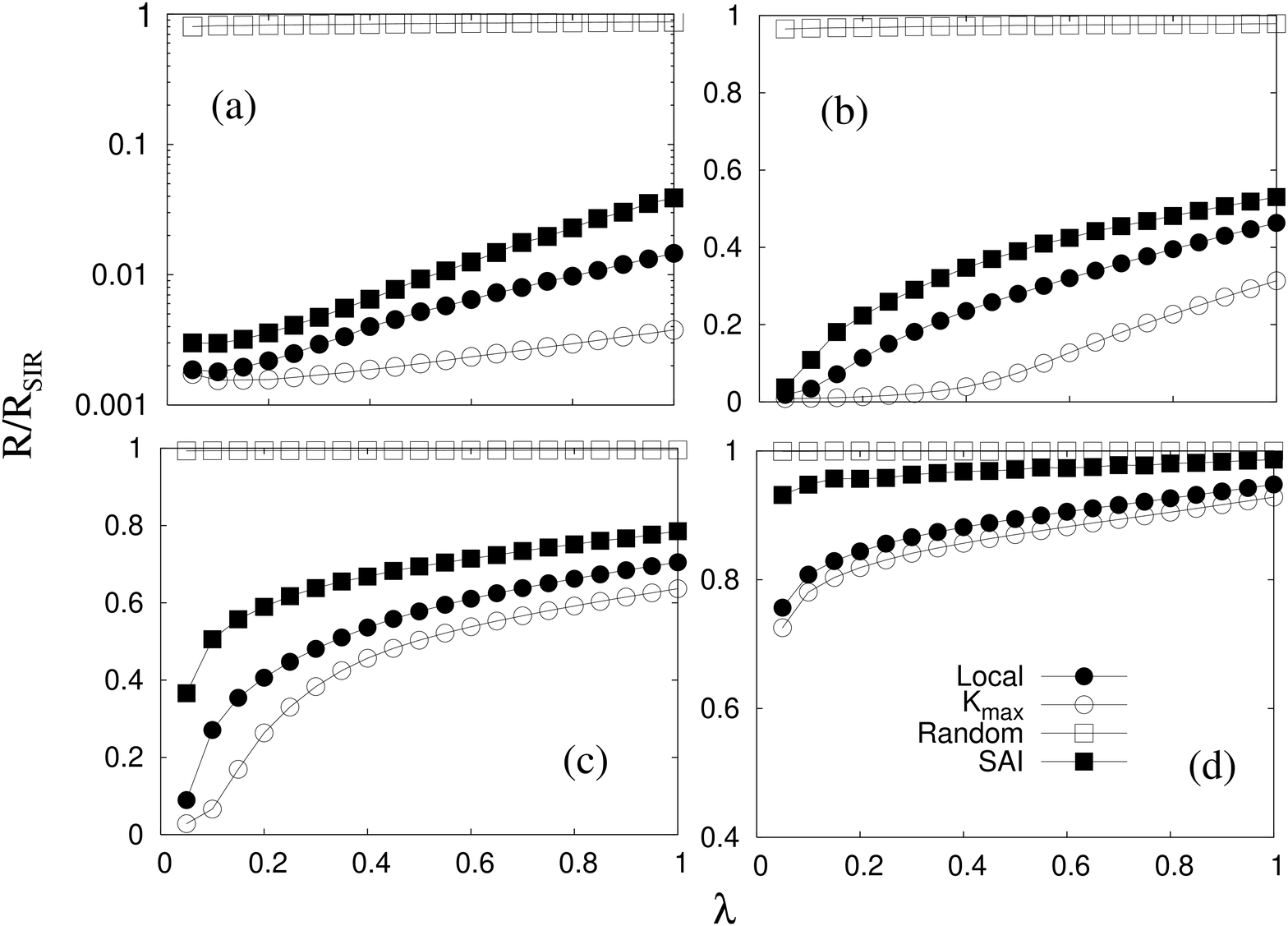,width=3.3in,angle=0,clip=1}
\end{center} 

\caption{Comparison of the immunization strategies for the Internet AS map. In
  the figure, we have represented the ratio between the epidemic incidence of
  the four immunization strategies considered $(R)$ and that of the original
  system without immunization ($R_{SIR}$) for different values of $\langle x
  \rangle$. The legend refers to the following immunization strategies: the one
  introduced in this paper (local), targeted immunization ($K_{max}$), random
  immunization (random) and single acquaintance immunization (SAI). In this
  case, 1\% of the non-immune nodes were initially infected at random. See the
  text for further details. The distances considered in the local algorithm
  are: (a) $d=1$, (b) $d=2$, (c) $d=3$,(d) $d=5$. }

\label{fig2} 
\end{figure}

In order to illustrate the importance of the local properties of the
network on the performance of the immunization, we analyze the results
when targeted immunization is implemented on each representation of
the Internet. In targeted immunization, a fraction of highly connected
nodes are immunized (i.e., do not get infected) in decreasing order of
their degrees. In the event that there are left $l$ immune nodes to be
distributed within a connectivity class $k$ containing $j>l$ nodes,
the $l$ immune nodes are randomly distributed within the $j$ nodes and
the results are averaged over at least 100 additional realizations of
this procedure. The results depicted in the figure suggest that the
degree-degree correlations is one of the main factors influencing the
performance of the immunization policy. We see that even for small
percentages of immune nodes, immunization performs better in the AS
graph. This may be due to the compact distribution of hubs (which play
a key role in targeted immunization) in the router map whereas for the
AS representation they are distributed throughout the whole
network. Therefore, in the AS representation, targeted immunization
works better because immune nodes are more efficient in cutting the
paths leading to poorly connected nodes, the more abundant. These
differences will become more apparent later on when local immunization
strategies come into play.

\section{Immunization Strategies}

Let us now summarize the local immunization strategy introduced in
this work.  The allocation of network resources to satisfy a given
service with the least use of resources, is a frequent problem in
communication networks. In our case, we would like to have a robust
system in front of a disease or virus spreading process but saving
resources, that is, using the minimum number of immune nodes. This is
a highly topical problem in communication networks as it might lead to
the developing and deploying of a digital immune system to prevent
technological networks from virus spreading. Recently \cite{pablo1},
we have studied a general covering problem in which every vertex is
either covered or has at least one covered node at a distance at most
$d$. In what follows, we show that the set of covered vertices $\cal
C$ can be taken as the set of nodes to be immunized.

\begin{figure}
\begin{center}

\epsfig{file=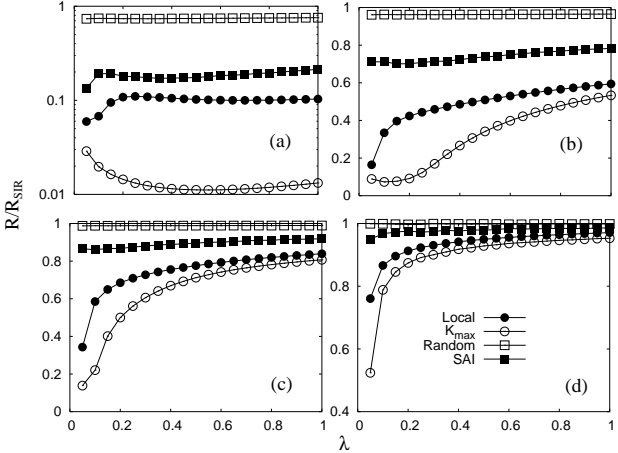,width=3.3in,angle=0,clip=1}

\end{center} 
\caption{Same as previous figure but for the Internet router map. The distances
  considered in the local algorithm are in this case: (a) $d=2$, (b) $d=5$, (c)
  $d=7$,(d) $d=10$.}
\label{fig3} 
\end{figure}

The heuristic algorithm proceeds as follows \cite{pablo1}: For every vertex $i$
in the network, look for the vertex with the highest degree within a distance
$d$ of $i$ and immunize it. In case there is more than one vertex with the
highest degree, one of them is selected at random and immunized. Moreover, if
there is already an immune vertex within the neighborhood of $i$, that
immunization is kept. We have shown before \cite{pablo1} that this local
algorithm gives near-optimal solutions for a general distance-$d$ covering
problem, though the result of the covering depends on topological features such
as the degree-degree correlations.

The immunization strategy here considered assumes that covered vertices are
immune nodes to the spreading of a disease or virus. For instance, in a
technological network, they could be thought of as being special devices
devoted to filtering out any virus or attack. This would imply that the
spreading process stops when it arrives to such nodes. This is of course the
ideal situation. However, it happens more often that immune nodes can not catch
the epidemic, but they are not able to stop spreading it through other nodes
$-$ as when you have an up-to-date anti-virus. Therefore, we study the worse
scenario and consider that immunized nodes just repel the virus cutting the
path to infection spreading.






The approach presented here is in the spirit of the immunization strategy
proposed by Cohen et al.\cite{cha04}. Since the immunization algorithm is
local, one only needs information about the neighbors of a given node up to a
distance $d$.  This information is usually available for small values of $d$
and easy to gather, in sharp contrast to targeted immunization that requires
complete knowledge of the degree distribution \cite{pv01c,db01a}. The
difference between our approach and that in \cite{cha04} is that we look for
the highly connected nodes in small parts of the network, while the strategy
developed in \cite{cha04} is based on the fact that randomly selected
acquaintances likely have larger connectivities than randomly chosen nodes.
Thus, in general, we expect our strategy to perform better than that proposed
in \cite{cha04}, while keeping the local character of the algorithm
\cite{note1}. On the other hand, either the number of immune nodes or the
distance $d$, which is a measure of the degree of local knowledge of the
network topology, should be fixed. This makes the algorithm more
parameter-constrained, but allows a more efficient distribution of resources.

We have performed extensive numerical simulations of four different
immunization schemes. The immunization obtained following the covering
algorithm fixes the fraction, $\langle x \rangle$, of immune nodes in
the whole network for each value of $d$. Random immunization means
that a fraction $\langle x \rangle$ of immune nodes is randomly placed
on the networks.  Targeted immunization looks for the $\langle x
\rangle N$ highly connected nodes and immunizes them. Finally, the
Single Acquaintance Immunization (SAI) algorithm proposed in
\cite{cha04} is run taking $p=\langle x \rangle$ and ensuring that the
total number of immune nodes is the same in both schemes.  In all
cases, the results are averaged over many realizations for each value
of $\lambda$ and $\langle x \rangle$. The results are displayed in
Fig.\ \ref{fig2} and Fig.\ \ref{fig3}.

As expected, targeted immunization produces the best results for both
topologies. Note that, as discussed in the previous section, the
performance of the algorithm depends on the specific topology and
produces different results for AS and router maps. On the other
extreme we find random immunization, whose performance is not affected
by the structure of the underlying networks. Turning our attention to
local algorithms, it is found that the immunization scheme based on
the covering algorithm performs better than the SAI, even for small
values of $d$, where it is truly local. In fact, it is outperformed
only by the targeted procedure and for all values of the parameters
$d$ and $\lambda$ it lies between the most efficient and the SAI
scheme. Additionally, from a practical point of view, the covering
strategy could be a good policy since it balances the degree of local
knowledge and the efficiency of the vaccination. Moreover, as all
network topologies are not neither completely known nor completely
unknown, the covering allows to fine-tune the value of $d$ on a
case-by-case base (that is, according to the degree of local knowledge
of the network) and thus it is more flexible than other immunization
strategies (recall that it is the result of an optimization).

We have further explored the differences between the global and
covering-based immunization schemes. In principle, one may think that
as we are immunizing highly connected nodes, both strategies produce
the same set of immune nodes.  Obviously, this is not the case since
the covering operates at shorter distances than targeted immunization
(which operates at $d=D$, the diameter of network). In fact, a direct
comparison of who the immune nodes are in both algorithms shows that
no more than $50\%$ of them are the same and both sets equal only when
$d$ reaches the diameter of the network.  Moreover, as a further
evidence of the influence of the graph representation in the
performance of immunization schemes, it is found that for the router
level the percentage above can increase up to $70\%$.

\begin{figure}[t]
\begin{center}

\epsfig{file=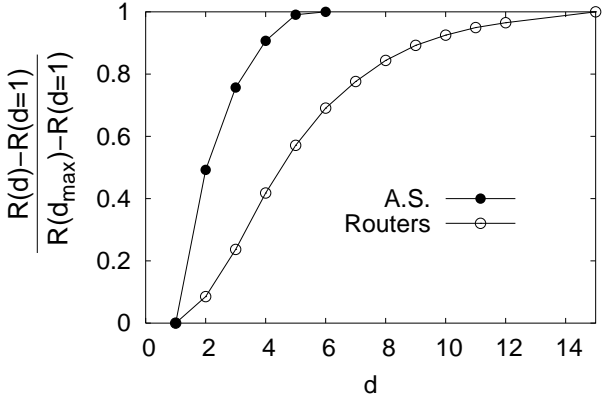,width=3.3in,angle=0,clip=1}

\end{center} 
\caption{Relative difference of the epidemic incidence for different
  values of $d$ with respect to that at $d=1$ ($\lambda=1$). The
  behavior observed in the figure is determined by the number of
  susceptible nodes each immune vertex has to ``protect''. See the
  text for further details.}
\label{fig5} 
\end{figure}

\begin{figure}
\begin{center}

\epsfig{file=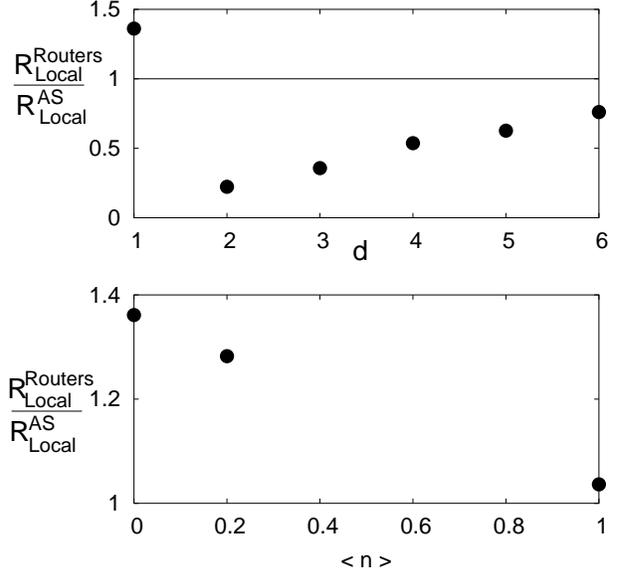,width=3.3in,angle=0,clip=1}

\end{center} 

\caption{Top: Phase transition is revealed by the best performance in
Routers when $\langle n\rangle$ is bigger for AS ($d>1$). Bottom: On
the contrary, when the nodes covered by each immune vertex 
$\langle n\rangle$ is (roughly) the same, immunization works better 
in A.S. ($\langle n\rangle=0$ for $d=1$; $\langle n\rangle\simeq0.2$
for $d=2$ in A.S. and $d=5$ in Routers;  $\langle n\rangle=1$ for
$d=6$ in A.S. and $d=15$ in Routers). The results were obtained starting from a
randomly chosen infected node and setting $\lambda=1$.}

\label{fig6} 
\end{figure}

Let us now restrict our discussion to the local (covering)
immunization scheme and focus on the influence of degree-degree
correlations on the final size of the outbreak. Figures\ \ref{fig5}
and \ref{fig6} reflect the differences in the algorithm's performance
for the AS and the Router maps of Internet. Figure\ \ref{fig5}
illustrates the relative difference of the epidemic incidence as a
function of $d$, taking as a reference the size of the outbreak at
$d=1$. The behavior depicted in the figure is quite similar to the
dependency of the number of nodes covered by each immune node,
$\langle n \rangle$, when $d$ is increased \cite{pablo1}. For the AS
network, the fraction of infected nodes at the end of the epidemic
spreading process rapidly increases. In contrast, the increase in the
epidemic incidence for the router network takes place at larger values
of $d$. This indicates that {\em for the same} $d>1$, the immunization
strategy works better at the router level as confirmed in Fig.\
\ref{fig6}, top panel.  The reason of this behavior becomes apparent
by noticing that for the router level $\langle x \rangle$ is bigger
than for the AS, but the number $\langle n \rangle$ of nodes
``covered'' on average by each immune node is smaller. The combination
of the two factors leads to a more efficient immunization at the
router level, however, at the cost of more resources. Both strategies
tend to be closer as $d$ is increased because at the router level the
correlations change beyond $d>2$.


The previous result has to be carefully interpreted and should not be
misunderstood. A closer look at the influence of the correlations
reveals that, although in general they determine $\langle x \rangle$
and $\langle n \rangle$ for each map, these two quantities {\em alone}
do not suffice to explain all the differences observed. Indeed, the
local structure of the network turns out to be at the root of the
immunization efficiency and the optimal trade-off between the size of
the outbreak and the least use of resources. To see this, we have
analyzed the situation in which {\em both} $\langle x \rangle$ (though
the $d$'s are different) and $\langle n \rangle$ are almost the same
in the two representations. This case is represented in the bottom
panel of Fig.\ \ref{fig6}. As can be seen from the figure, in the
latter case, the immunization scheme for the AS outperforms that for
the router level. This behavior is due to the fact that in the AS
network, the immune nodes are more distributed throughout the network
because highly connected vertices alternate with poorly connected
ones. On the contrary, at the router level, the hubs are topologically
closer to each other (the correlations are positive) and thus some of
the immune nodes are not highly connected resulting in a less
efficient protection in front of an epidemic.

\section{Discussion and Conclusions}

In this paper, we have analyzed the spreading of an epidemic disease
on top of real communication networks both with and without
immunization. First, we have shown that targeted immunization produces
different results depending on the local properties of the underlying
graph by using different representations of the same technological
network, the Internet. Later, we turned our attention to several
immunization strategies and proposed a scheme that is neither
completely local nor global, but can be tuned between the two
extremes. The strategy introduced has been shown to perform better
than all previous methods irrespective of the degree of local
knowledge, except for the case of targeted immunization.

An important part of the work has dealt with the influence of
degree-degree correlations on the performance of all vaccination
algorithms. To this respect, it has been shown that local properties
are extremely important for the outcome of a given strategy. Moreover,
the work presented here has been performed on top of real networks, an
thus the results are of high practical interest. An added value of the
method developed here is that the covering-based strategy does not
only deal with the degree of the immune nodes, as targeted
immunization does, but naturally introduces the practical constraint
of having limited resources to be distributed in the system on top of
which the epidemics is spreading.  Therefore, our method and the
results found can shed light and provide useful hints in the search of
optimal immunization strategies as the development and deploying of a
digital immune system, a highly topical issue nowadays.

Finally, it is worth mentioning that although we have not analyzed the
case here, it would also be possible to develop an even more flexible
strategy in which the immunization through the covering algorithm is
done with a variable $d$ for the same network, that is, one can
implement an algorithm that optimize $\langle x \rangle$ locally for
different neighborhoods (i.e., different values of $d$ for each
neighborhood) of a given (large) network.

In summary, our work points to a new direction in designing
immunization strategies, namely, the finding of a better trade-off
between resources and algorithm's performance.

\section{Acknowledgments}
  We thank A.\ V\'azquez for discussions and for providing the data
  used in this work. P.\ E.\ and J.\ G-G\ acknowledge financial
  support of the MEC through FPU grants. Y.\ M.\ is supported by MEC
  through the Ram\'{o}n y Cajal Program.  This work has been partially
  supported by the Spanish DGICYT Projects FIS2004-05073-C04-01,
  BFM2002-00113 and BFM2003-08532; and a DGA (Arag\'on, Spain)
  project.


\begin{thebibliography}{99}
\bibitem{note1} Strictly speaking, our algorithm is neither completely
  local nor global. In fact, by tuning the distance $d$ of the
  immunization (covering) strategy one can move from a truly local
  algorithm to an algorithm close to the targeted immunization
  approach for $d\sim D$, being $D$ the diameter of the network. In
  this sense, our method is half-a-way between strictly local and
  global strategies. This difference diffuses when one consider
  ultra-small world networks, which is not our case.
\end{thebibliography}
\end{document}